\documentclass[%
 reprint,
 amsmath,amssymb,
 aps,
]{revtex4-1}

\usepackage{graphicx}
\usepackage{dcolumn}
\usepackage{bm}

\begin{document}

\preprint{APS/123-QED}

\title{Steering in spin tomographic probability representation}

\author{V.I. Man'ko}
 \email{manko@na.infn.it}
\affiliation{P.N. Lebedev Physical Institute, Russian Academy of Sciences
Leninskii Prospect 53,\\ Moscow 119991, Russia}
\author{L.A. Markovich}%
 \email{kimo1@mail.ru}
\affiliation{
 Moscow Institute of Physics and Technology (State University)
Institutskii per. 9, Dolgoprudnyi, Moscow Region 141700, Russia
}%

\date{\today}

\begin{abstract}
The steering property known for two qubit-state in terms of specific inequalities for correlation function is translated for the state of qudit with the spin $j=3/2$.
Since most steering detection inequalities are based on the correlation functions we introduce analogs of such functions for the single qudit systems. The tomographic probability representation for the qudit states is applied. The connection between the correlation function in two qubit system and the single qudit is  presented in an integral form with a intertwining kernel calculated explicitly in tomographic probability terms.
\begin{description}
\item[PACS numbers]
03.65.Ud, 03.67.Mn
\end{description}
\end{abstract}

\pacs{03.65.Ud}
\pacs{03.67.Mn}                           
\keywords{Steering, tomography, qudit system}
\maketitle


\section{\label{sec:level1}Introduction}
The problem of the quantum steering was introduced by E. Schr\"{o}dinger in \cite{Schrodinger} as an answer to the paper of A. Einstein, B. Podolsky and N. Rosen \cite{Einstein} to generalize the EPR paradox. Since  the EPR steering can be applied in one-way quantum cryptography \cite{Chen,Saunders} or in visualization of the two-qubit state tomography \cite{Jevtic} the problem is discussed in a large number of recent papers.
\par There are many definitions of the term 'steering'. In \cite{Marciniak} the EPR steering was defined as a form of a nonlocality in quantum mechanics, that is in between the entanglement and the Bell nonlocality. In \cite{Wiseman} the notion of steering was reformulated. The EPR steering was considered as the ability of the first system to affect the state of the second system through the choice of the first systems measurement basis.  Hence, the concept of the quantum steering can be introduced not only for the multipartite (joint) systems, but also for all systems (including non composite ones)  with correlations \cite{Mar_8}.
\par The EPR steering can be detected through the violation of steering inequalities \cite{Saunders,Zukowski,Schneeloch,Schneeloch2}. Most of the steering inequalities are
connected with the notion of the correlation function \cite{Zukowski}. Therefore the main focus of our paper is the representation of the correlation function for the two-qubit and for the single qudit systems. To characterize degrees of quantum correlations in the systems we use the tomographic probability representation of quantum mechanics \cite{Ibort}.
We find the tomographic representation of the correlation functions that characterize the steering in the qunatum system. We introduce the connection between tomograms for the two-qubit system and the tomogram for the single qudit with the spin $j=3/2$. To introduce the correlation function we consider the specific observable one which is a complete analog of the observable used in two qubit system but studied in the single qudit with the spin $j=3/2$ picture.
 Hence, we can introduce the notion of the steering and detect the steering phenomenon in the system without subsystems. The physical application of the steering phenomena in the single qudit with the spin $j=3/2$ and for the four level atom can be  performed in the study of the information and entropic properties of the  superconducting multilevel circuits \cite{Kiktenko,Glushkov}.
\par The paper is organized as follows. In  Sec. \ref{sec:1} we rewrite the correlation function for the two-qubit system in terms of the spin tomogram. The connection of the two partite system tomogram and the single qudit state tomogram is introduced in Sec. \ref{sec:2}.
In  Sec. \ref{sec:3} the correlation function is rewritten by means of the tomogram and the notion of the EPR steering is extended to the case of the system without subsystems (a single qudit).
\section{The Two-qubit Steering in the Spin Tomographic Representation}\label{sec:1}
The notion of the EPR steering is best studied on the example of the two-qubit state, where each qubit is defined on two dimensional Hilbert space.
Let us define the two-qubit quantum system on the Hilbert space. The density matrix of the system state in a four-dimensional Hilbert space $\mathcal{H}_{AB}$ is the matrix $\rho_{AB}$ of the size $4\times4$ with the nonnegative eigenvalues and $\rho_{AB}=\rho_{AB}^{\dagger}$, $Tr\rho_{AB}=1$ hold.
The correlations in the such system $\rho_{AB}$ can be described by the joint probability function
 \begin{eqnarray}\label{2}P(a,b|A, B)&=&\int p_{\lambda}P(a|A,\lambda)P(b|B,\lambda)d\lambda,
     \end{eqnarray}
     where $P(a|A,\lambda)$ is the probability distribution of  the measurement outcomes $a$ under setting $A$ for a hidden variable $\lambda$. The hidden variable has the probability distribution $p_{\lambda}$ and its hidden state is $\rho_{\lambda}$ (a local hidden state (LHS)). If the following model of the correlation
      \begin{eqnarray}P(a,b|A, B)&=&\int p_{\lambda}P(a|A,\lambda)Tr(\widehat{\pi}(b|B)\rho^{(b)}_{\lambda})d\lambda
     \end{eqnarray}
     does not exist, then the state is steerable \cite{Zukowski}. Here  $\widehat{\pi}(b|B)$ is the projection operator for an observable parameterized by the setting $B$ and the $\rho^{(b)}_{\lambda}$ is some pure state of the system $B$. The EPR steering can be detected through the violation of the steering inequalities. The steering inequalities are mostly based on the notion of the correlation function. The quantum correlation function for the two-qubit state is determined by
 \begin{eqnarray} \label{4}E(\overrightarrow{k_1},\overrightarrow{k_2})&=&Tr(\overrightarrow{k_1}\cdot\overrightarrow{\sigma}\otimes\overrightarrow{k_2}\cdot\overrightarrow{\sigma}\rho),
    \end{eqnarray}
 where  $\overrightarrow{\sigma}$ is the vector built out of the Pauli matrix, $\overrightarrow{k_1}$, $\overrightarrow{k_2}$ are the unit Bloch vectors of the measurement directions equal to $\pm1$.
\par The tomographic probability distribution of the spin states allows to describe the states determined by the density matrix $\rho $ of the two qubits by means of the tomogram.
By definition the spin tomogram
\begin{eqnarray*}\omega(\mathbf{x})=\omega(m_1,m_2,u)&=&<m_1m_2|u\rho u^{\dag}|m_1m_2>\end{eqnarray*}
 is the probability to obtain $m_{1}=-j_1,-j_1+1,\ldots,j_1$, $m_{2}=-j_2,-j_2+1,\ldots,j_2$, $j_{1,2}= 0,1/2,1\ldots$  as the spin projections on directions given by the unitary matrix $u$. Here we used the notation $\mathbf{x}=(m_1,m_2,u)$ and  the matrix $u$ is the rotation matrix of the size $N\times N$, where  $N=(2j_1+1)(2j_2+1)$ holds. If we choose the rotation matrix as the direct product of the two matrices $u=u_1\otimes u_2$ of irreducible representations of the $SU(2)$ - group, i.e.
 \begin{eqnarray*}u_j&=&\left(
                       \begin{array}{cc}
                         \cos\frac{\theta_j}{2} e^{\frac{i(\varphi_j+\psi_j)}{2}}& \sin\frac{\theta_j}{2} e^{\frac{i(\varphi_j-\psi_j)}{2}} \\
                         -\sin\frac{\theta_j}{2}e^{\frac{i(\psi_j-\varphi_j)}{2}} & \cos\frac{\theta_j}{2} e^{\frac{-i(\varphi_j+\psi_j)}{2}}\\
                       \end{array}
                     \right),\quad j=1,2
\end{eqnarray*}
  then the latter tomography can be written as
\begin{eqnarray}&&\omega(m_1,m_2,u_1,u_2)=\\
&=&<m_1m_2|u_1\otimes u_2\rho u_1^{\dag}\otimes u_2^{\dag}|m_1m_2>.\label{16}\nonumber\end{eqnarray}
The matrices $u_1$ and $u_2$ depend only on the Euler angles $\{\theta_i,\varphi_i,\psi_i\},i=\{1,2\}$ which determine the directions of quantization, e.g., points on the Bloch sphere. Hence, we use the following notations $u_1=u_1(\theta_1,\varphi_1,\psi_1)=u_1(\vec{n}_1)$, $u_2=u_2(\theta_2,\varphi_2,\psi_2)=u_2(\vec{n}_2)$, where $\vec{n}_1$ and $\vec{n}_2$ determine directions of spin projection axes. Hence, tomogram \eqref{16} can be rewritten as
\begin{eqnarray}\label{1}&&\omega(m_1,m_2|\vec{n}_1,\vec{n}_2)=\\
&=&<m_1m_2|u_1(\vec{n}_1)\otimes u_2(\vec{n}_2)\rho u_1^{\dag}(\vec{n}_1)\otimes u_2^{\dag}(\vec{n}_2)|m_1m_2>.\nonumber\end{eqnarray}
The latter tomogram is the conditional probability of projections of spins $m_1$, $m_2$ on vectors $\vec{n}_1$, $\vec{n}_2$ on the Bloch sphere. Hence, \eqref{1} is the tomographic representation of the probability
\eqref{2}. The probability function \eqref{1} has the property of a no-signaling. Hence, the marginal probability distributions of the first and the second qubit are
\begin{eqnarray}\label{3}\omega_1(m_1|\vec{n}_1)&=&\sum\limits_{m_2=-j_2}^{j_2}\omega_1(m_1,m_2|\vec{n}_1),\\
\omega_1(m_2|\vec{n}_2)&=&\sum\limits_{m_1=-j_1}^{j_1}\omega_1(m_1,m_2|\vec{n}_2)\nonumber.
\end{eqnarray}
It is known that the state is called separable if and only if the density operator of the composite system $\rho$
can be written as
\begin{eqnarray*}\rho&=&\sum\limits_{k}p_k\rho_1^{(k)}\otimes\rho_2^{(k)},\quad \sum\limits_{k}p_k=1.\end{eqnarray*}
Hence, using \eqref{3} tomogram \eqref{1} can be written as
\begin{eqnarray*}&&\omega(m_1,m_2|\vec{n}_1,\vec{n}_2)=\sum\limits_{k}p_k\omega_1^{(k)}(m_1|\vec{n}_1)\omega_2^{(k)}(m_2|\vec{n}_2)\\
&=&\sum\limits_{\lambda}p(\lambda)\omega_1(m_1|\vec{n}_1,\lambda)\omega_2(m_2|\vec{n}_2,\lambda).\end{eqnarray*}
We can rewrite the latter expression in the form \eqref{2}, which is the LHS model in the form of the tomogram.
\par The tomogram can be represented using the operator $\widehat{U}(m_1, m_2, \vec{n}_1,\vec{n}_2)$ called the dequantizer. What's more, by the given spin tomogram one can reconstruct the operator of the density matrix $\widehat{\rho}$ using the operator $\widehat{D}(m_1, m_2, \vec{n}_1,\vec{n}_2)$ called the quantizer
\begin{eqnarray}\label{13}&&\omega(m_1, m_2, \vec{n}_1,\vec{n}_2)=Tr(\widehat{\rho}\widehat{U}(m_1, m_2, \vec{n}_1,\vec{n}_2)),\\\nonumber
&&\widehat{\rho}=\sum\limits_{m_1,m_2}\int\omega(m_1, m_2, \vec{n}_1,\vec{n}_2)\widehat{D}(m_1, m_2, \vec{n}_1,\vec{n}_2)d\vec{n}_1d\vec{n}_2,
\end{eqnarray}
where it holds
\begin{eqnarray*}\int d\vec{n}&=&\int\limits_{0}^{2\pi}d\varphi\int\limits_{0}^{\pi}\sin\theta d\theta\int\limits_{0}^{2\pi}d\psi.
\end{eqnarray*}
In \cite{Filippov} the dequantizer and the quantizer operators for the two-qubit state are defined as
\begin{eqnarray}\label{8}&&\widehat{U}(m_1, m_2, \vec{n}_1,\vec{n}_2)=\widehat{U}(m_1,\vec{n}_1)\otimes\widehat{U}(m_2,\vec{n}_2)\\
&=&\left(\frac{1}{2}\widehat{I}+m_1 F(\varphi_1, \theta_1)\right)\otimes \left(\frac{1}{2}\widehat{I}+m_2 F(\varphi_2, \theta_2)\right),\nonumber\\
&&\widehat{D}(m_1, m_2, \vec{n}_1,\vec{n}_2)=\widehat{D}(m_1,\vec{n}_1)\otimes \widehat{D}(m_2,\vec{n}_2)\nonumber\\
&=&\left(\frac{1}{8\pi^2}\left(\frac{1}{2}\widehat{I}+3m_1 F(\varphi_1, \theta_1)\right)\right)\otimes\\
 &\otimes&\left(\frac{1}{8\pi^2}\left(\frac{1}{2}\widehat{I}+3m_2 F(\varphi_2, \nonumber \theta_2)\right)\right),
\end{eqnarray}
where $\widehat{I}$ is the $2\times2$ identity matrix and
\begin{eqnarray*}F(\varphi, \theta)=\left(
                                      \begin{array}{cc}
                                        \cos\theta & -e^{i\varphi}\sin\theta \\
                                       -e^{-i\varphi}\sin\theta & -\cos\theta \\
                                      \end{array}
                                    \right).
\end{eqnarray*}
\par Any observable $A$ can be identified with the a Hermitian operator $\widehat{A}$. In \cite{Marmo25} the tomographic symbol $\omega_{A}(\mathbf{x})$ of the operator $\widehat{A}$ is determined by
      \begin{eqnarray*}\omega_{A}(\mathbf{x})=Tr(\widehat{A}\widehat{U}(\mathbf{x})),\quad \widehat{A}=\int\omega_{A}(\mathbf{x})\widehat{D}(\mathbf{x})d\mathbf{x},\end{eqnarray*}
      where $\widehat{U}(\mathbf{x})$, $\widehat{D}(\mathbf{x})$ are the dequantizer and quantizer operators, respectively.
Using the quantizer operator as the dequantizer and vise versa the dual tomographic symbol $\omega_{A}^d(\mathbf{x})$ is introduced
      \begin{eqnarray}&&\omega_{A}^d(\mathbf{x})=Tr(\widehat{A}\widehat{U}'(\mathbf{x}))=Tr(\widehat{A}\widehat{D}(\mathbf{x})),\\\nonumber &&\widehat{A}=\int\omega_{A}^d(\mathbf{x})\widehat{D}'(\mathbf{x})d\mathbf{x}=\int\omega_{A}^d(\mathbf{x})\widehat{U}(\mathbf{x})d\mathbf{x}.\label{17}
   \end{eqnarray}
   Using the dual tomographic symbol \eqref{17} we can write the following trace
   \begin{eqnarray}\label{55}Tr(\widehat{A}\widehat{\rho})=\int\omega_{A}(\mathbf{x})\omega_{\rho}^d(\mathbf{x})d\mathbf{x}.
   \end{eqnarray}
   Hence, using \eqref{55} the quantum correlation function \eqref{4} can be rewritten in the tomographic form
         \begin{eqnarray}\label{14}E(\overrightarrow{k_1},\overrightarrow{k_2})=\int\omega_{B}(\mathbf{x})
         \omega_{\rho}^d(\mathbf{x})d\mathbf{x}
   \end{eqnarray}
   or in the equivalent form
            \begin{eqnarray}\label{14_2}&&E(\overrightarrow{k_1},\overrightarrow{k_2})=\int\omega_{\rho}(\mathbf{x})\omega^d_{B}(\mathbf{x})
         d\mathbf{x}\\
         &=&\sum_{m_1,m_2}\int \omega_{\rho}(m_1, m_2, \vec{n}_1,\vec{n}_2)
         \omega^d_{B}(m_1, m_2, \vec{n}_1,\vec{n}_2)
         d\vec{n}_1d\vec{n}_2,\nonumber
   \end{eqnarray}
   where we used the notation $B=k_1\sigma\otimes k_2\sigma$.
   Thus, we can write any steering inequity that contains the correlation function in terms of the spin tomograms.
   \section{The Relation Between the Single Qudit and the Two-qubit States Tomograms}\label{sec:2}
   For the single qudit with the spin $j=3/2$ and the density matrix $\rho$ we can write the tomographic representation as
      \begin{eqnarray}\label{6}&&\widehat{\rho}=\sum\limits_{m=-3/2}^{3/2}\int W(m,\vec{n})\widehat{\mathcal{D}}(m,\vec{n}) d\vec{n},\\ &&W(m,\vec{n})=Tr\left(\widehat{\mathcal{U}}(m,\vec{n})\widehat{\rho}\right),\nonumber
   \end{eqnarray}
   where $\widehat{\mathcal{U}}(m,\vec{n})$ and $\widehat{\mathcal{D}}(m,\vec{n})$ are the dequantizer and the quantizer operators of the latter state, respectively. The tomogram $W(m,\vec{n})$
is the conditional probability of projections of the spin $m$ on vector $\vec{n}$ on the Bloch sphere. Note that the later tomogram depends on less amount of numbers comparable to the two-qubit state tomogram \eqref{1}.
   \par In \cite{MankoJetp} it was obtained that any qudit state with the spin $j$ can be represented  as
   \begin{eqnarray*}&&\widehat{\rho}^{(j)}=\sum\limits_{m=-j}^{j}\int \frac{d\omega}{8\pi^2}W(m,\alpha,\beta)\widehat{B}^{j}_{m}(\alpha,\beta),
   \end{eqnarray*}
   where $\widehat{B}^{j}_{m_1}(\alpha,\beta)$ is the quantizer operator. For the single qudit state, the latter operator is defined by the following matrix
    \begin{eqnarray*}  \label{7}B^{3/2}_{m}(\alpha,\beta)&=&B_{1m}(\alpha,\beta)+ \frac{i(-1)^{m}}{2(m + \frac{3}{2})!(\frac{3}{2} - m)!}\\
    &\cdot&\left(5mB_{2m}(\alpha,\beta)+\frac{21}{2}\sin \beta B_{3m}(\alpha,\beta)\right),\nonumber\end{eqnarray*}
   where we used the notations
   \begin{widetext}
\begin{equation*}
{B_{1m}(\alpha,\beta)}=\left(
                             \begin{array}{cccc}
                               \frac{1}{4}+ \frac{9}{10 }m\cos \beta &  \frac{3\sqrt{3}m}{10}\sin \beta e^{-\alpha i}& 0 & 0 \\
                               \frac{3\sqrt{3}m}{10}\sin \beta e^{\alpha i}& \frac{1}{4}+ \frac{3}{10}m\cos \beta & \frac{63m}{105}\sin \beta e^{\alpha i} & 0 \\
                               0 &\frac{63m}{105}\sin \beta e^{\alpha i} & \frac{1}{4} - \frac{3}{10}m\cos \beta & \frac{3\sqrt{3}m}{10}\sin \beta e^{-\alpha i} \\
                               0 & 0 & \frac{3\sqrt{3}m}{10}\sin \beta e^{\alpha i} & \frac{1}{4}- \frac{9}{10}m\cos \beta \\
                             \end{array}
                           \right)
 \;,
\end{equation*}
\begin{equation*}
{B_{2m}(\alpha,\beta)}=\left(\begin{array}{cccc}
                                   3\cos^2 \beta - 1 & \sqrt{3}e^{-\alpha i}\sin 2\beta &\sqrt{3}\sin^2\beta e^{-2\alpha i}  & 0 \\
                                   \sqrt{3}\sin 2\beta e^{ai}&  - 3\cos^2 \beta+ 1& 0 &  -\sqrt{3}\sin^2\beta e^{-2\alpha i}\\
                                    \sqrt{3}\sin^2\beta e^{2ai} & 0 &  -3\cos^2\beta + 1 & -\sqrt{3}\sin 2\beta e^{-\alpha i} \\
                                     0& -\sqrt{3}\sin^2\beta e^{2ai}& -\sqrt{3}\sin 2\beta e^{ai}&  3\cos^2\beta - 1 \\
                                  \end{array}\right)
 \;,
\end{equation*}
\begin{equation*}
{B_{3m}(\alpha,\beta)}=
\left(
                      \begin{array}{cccc}
                  \frac{\cos \beta}{\sin \beta}(\cos^2\beta -\frac{3}{5}) & \sqrt{3}\left(\cos^2\beta -\frac{1}{5}\right)e^{-\alpha i} &\sqrt{3}\sin\beta\cos \beta e^{-2\alpha i} &
                                  \sin^2\beta e^{-3\alpha i} \\
                      \sqrt{3}\left(\cos^2\beta -\frac{1}{5}\right) e^{\alpha i}  & 3\frac{\cos \beta}{\sin \beta}(\cos^2\beta - \frac{3}{5})&3(\frac{1}{5}  -\cos^2\beta)e^{-\alpha i} &
                                  -\sqrt{3}\sin\beta \cos \beta e^{-2\alpha i} \\
                                 \sqrt{3}\sin\beta \cos \beta e^{2\alpha i} &
                                 3(\frac{1}{5} - \cos^2\beta)e^{\alpha i}&
                                  - 3\frac{\cos \beta}{\sin \beta}(\cos^2\beta- \frac{3}{5})&
                                  \sqrt{3}\left(\cos^2\beta -\frac{1}{5}\right)e^{-\alpha i} \\
                       \sin^2 \beta e^{3\alpha i} &
                                        \sqrt{3}\sin\beta\cos \beta e^{2\alpha i}&
                                         \sqrt{3}(\cos^2\beta - \frac{1}{5})e^{\alpha i}&
                                          -\frac{\cos \beta}{\sin \beta}(\cos^2\beta - \frac{3}{5})\\
                      \end{array}
                    \right)
\;.
\end{equation*}
\end{widetext}
   We can use the latter operator as the quantizer for the single qudit system, i.e $\widehat{\mathcal{D}}(m,\alpha,\beta)=\frac{1}{8\pi^2}\widehat{B}^{3/2}_{m}(\alpha,\beta)$.
   To write the rotation matrix $U^{(3/2)}(\alpha,\beta,\gamma)$ for the single qudit state  we can use the Wigner's D-function
   \begin{eqnarray*}D^{(j)}_{m',m}(\alpha,\beta,\gamma)&=&e^{im'\gamma}d^{(j)}_{m',m}(\beta)e^{im'\alpha},
   \end{eqnarray*}
   where it holds
   \begin{eqnarray}\label{12}
d_{m',m}^{(j)}(\beta)&=&\left(\frac{(j+m')!(j-m')!}{(j+m)!(j-m)!}\right)^{1/2}\cos\left(\frac{\beta}{2}\right)^{m'+m}\\
&\cdot&\sin\left(\frac{\beta}{2}\right)^{m'-m}
P_{j-m'}^{(m'-m,m'+m)}(\cos\beta)\nonumber,
\end{eqnarray}
and $P_{j-m'}^{(m'-m,m'+m)}(\cos\beta)$ denote the Jacobi polynomials \cite{Landau}.
 Hence, the rotation matrix for the single qudit state $U^{(3/2)}(\alpha,\beta,\gamma)$ has elements $U^{(3/2)}_{m',m}(\alpha,\beta,\gamma)=D^{(3/2)}_{m',m}(\alpha,\beta,\gamma)$.
Using the latter rotation matrix the dequntizer operator can be defined by the following matrices
      \begin{eqnarray*}\mathcal{U}(m,\alpha,\beta)&=&U^{(3/2)\dag}(\alpha,\beta,\gamma)|m><m|U^{(3/2)}(\alpha,\beta,\gamma).
   \end{eqnarray*}
To find the relation between the two-qubit system  and the single qudit system tomograms we substitute \eqref{6} into \eqref{13}. We get
      \begin{eqnarray}\label{11}&&\omega(m_1, m_2, \vec{n}_1,\vec{n}_2)=\\
       &=&\int Tr W(m,\vec{n}) \widehat{\mathcal{D}}(m,\vec{n}) \widehat{U}(m_1, m_2, \vec{n}_1,\vec{n}_2)d\vec{n}\nonumber\\
      &=&\int W(m,\vec{n})K_{12}d\vec{n}\nonumber,
   \end{eqnarray}
   where  the notation $K_{12}\equiv K_{12}(m_1, m_2,m,\vec{n}_1,\vec{n}_2, \vec{n})$ $= Tr  \widehat{\mathcal{D}}(m,\vec{n}) \widehat{U}(m_1, m_2, \vec{n}_1,\vec{n}_2)$ is introduced. We call it the kernel function. Using \eqref{8} and \eqref{7} we can write the latter kernel in the explicit form
      \begin{widetext}
       \begin{eqnarray}\label{9}
       &&{K_{12}(m,m_1,m_2,\alpha,\beta,\theta_1,\theta_2,\varphi_1,\varphi_2)}=\frac{1}{4}+\frac{3m}{5}\Bigg(\cos\beta(2m_1\cos\theta_1+m_2\cos\theta_2)\\\nonumber
       &+&m_2\sin\beta\cos\alpha\sin\theta_2e^{i\varphi_2}\left(-\sqrt{3}+2m_1 \sin\theta_1e^{i\varphi_1}\right)\Bigg)
       -\frac{(-1)^mi}{(m + 3/2)!(3/2 - m)!}\\\nonumber
       &\cdot&\Bigg(21\cos\beta\left(-\frac{3}{5}+cos^2\beta\right)\left(\frac{m_1}{2}\cos\theta_1-m_2\cos\theta_2\right)+10m m_1m_2\cos\theta_1\cos\theta_2\left(1-3\cos^2\beta\right)\\\nonumber
       &+&\sqrt{3}m_2\Bigg(\frac{21}{2}\sin\beta \sin\theta_2 e^{i\varphi_2}\cos\alpha\left(\cos^2\beta-\frac{1}{5}\right)
       +21m_1\cos\beta\sin^2\beta\cos\theta_2\sin\theta_1e^{i\varphi_1}\cos2\alpha\\\nonumber
       &+&10m m_1\sin^2\beta\Bigg(e^{i\varphi_1}\cos\theta_2\sin\theta_1\cos2\alpha+
      4e^{i\varphi_2}\cos\theta_1\sin\theta_2\cos\alpha\Bigg)\Bigg)\\\nonumber
      &+&\frac{21m_1m_2}{2}\sin\beta\sin\theta_1\sin\theta_2e^{i\varphi_1}e^{i\varphi_2}\Bigg(-\frac{3}{5}\cos\alpha
      +3\cos^2\beta\cos\alpha-\sin^2\beta\cos3\alpha
      \Bigg)
   \end{eqnarray}
      \end{widetext}
   The kernel depends on three quantum numbers and six angles.
     Similarly we can write the inverse transformation as
           \begin{eqnarray*}W(m,\vec{n})=\int \omega(m_1, m_2, \vec{n}_1,\vec{n}_2) K_{21}d\vec{n_1}d\vec{n_2},
   \end{eqnarray*}
   where the kernel is $K_{21}\equiv K_{21}(m_1, m_2,m,\vec{n}_1,\vec{n}_2, \vec{n})=Tr \widehat{D}(m_1, m_2, \vec{n}_1,\vec{n}_2)\widehat{\mathcal{U}}(m,\vec{n}) $.
   For the dual tomographic symbols \eqref{17} it is easy to see that
              \begin{eqnarray*}&&\omega^{d}(m_1, m_2, \vec{n}_1,\vec{n}_2)=\int W^{d}(m,\vec{n})K^d_{12}d\vec{n},\\
              &&W^{d}(m,\vec{n}) =\int \omega^{d}(m_1, m_2, \vec{n}_1,\vec{n}_2) K^{d}_{21}d\vec{n_1}d\vec{n_2},
   \end{eqnarray*}
   where the kernels are $K^d_{12}=K_{21}$ and $K^d_{21}=K_{12}$.
   The latter four kernels describe the connection between the tomograms and the dual tomographic symbols for the bipartite and single qudit states.
      \subsection{The Example}
      Let us have the quantum state described by the $4\times4$ Werner density matrix
      \begin{eqnarray}\label{10_1}\rho^{W}&=&\left(
                     \begin{array}{cccc}
                       \frac{1+p}{4} & 0 & 0 & \frac{p}{2}\\
                       0 & \frac{1-p}{4}& 0 & 0 \\
                       0 & 0& \frac{1-p}{4} & 0 \\
                       \frac{p}{2} & 0 & 0 & \frac{1+p}{4} \\
                     \end{array}
                   \right),
\end{eqnarray}
where the parameter $p$ satisfies the inequality $-\frac{1}{3}\leq p\leq1$. The parameter domain $\frac{1}{3}< p\leq1$ corresponds to the entangled state.
If matrix \eqref{10_1} describes the single qudit state with the spin $j=3/2$ the tomogram \eqref{6} has the following elements
         \begin{widetext}
   \begin{eqnarray*}W\left(-\frac{3}{2},\alpha,\beta\right)& =& \frac{p}{16}+\frac{3p}{16}\cos2\beta -\frac{3p}{32}\sin\beta \cos3\alpha+\frac{p}{32}\cos3\alpha\sin3\beta+\frac{1}{4}\\
 W\left(-\frac{1}{2},\alpha,\beta\right)&=& \frac{3p}{16}(2\sin^2\beta-1)-\frac{p}{16}+\frac{3p}{32}\sin3\beta(2\sin\frac{3\alpha}{2}^2-1)-\frac{9p}{32}\sin\beta(2\sin\frac{3\alpha}{2}^2-1)+\frac{1}{4}\\
W\left(\frac{1}{2},\alpha,\beta\right)&=&  \frac{3p}{16}(2\sin^2\beta-1)-\frac{p}{16}-\frac{3p}{32}\sin3\beta(2\sin\frac{3\alpha}{2}^2-1)+\frac{9p}{32}\sin\beta(2\sin\frac{3\alpha}{2}^2-1)+\frac{1}{4}\\
W\left(\frac{3}{2},\alpha,\beta\right)&=& \frac{p}{16}+ \frac{3p}{16}\cos2\beta +\frac{3p}{32}\sin\beta \cos3\alpha-\frac{p}{32}\cos3\alpha\sin3\beta+\frac{1}{4}.
   \end{eqnarray*}
         \end{widetext}
From the other hand, the density matrix \eqref{10_1} can describe the two-qubit state. Hence, substituting the latter tomogram and the kernel \eqref{9} into \eqref{1} we can get the tomogram for the two-qubit Werner density matrix
         \begin{widetext}
\begin{eqnarray*}\omega(m_1, m_2, \theta_1,\theta_2,\phi_1,\phi_2)&=&2\pi\sum_{m=-3/2}^{3/2}\int_{0}^{2\pi}\int_{0}^{\pi}\sin\beta W(m,\alpha,\beta)K_{12}(m,m_1, m_2, \alpha,\beta, \theta_1,\theta_2,\phi_1,\phi_2)d\beta d\alpha\\
&=&\frac{1}{4}+pm_1m_2\left(\cos\theta_1\cos\theta_2+\sin\theta_1\sin\theta_2e^{i\phi_1}e^{i\phi_2}\right)
   \end{eqnarray*}
            \end{widetext}
which coincides with \eqref{13}.
      \section{The Single Qudit Steering in the Spin Tomographic Representation}\label{sec:3}
      In Sec. \ref{sec:1} the correlation function $E(\overrightarrow{k}_1,\overrightarrow{k}_2)$ is obtained by measuring the quantum observable $\widehat{O}=\overrightarrow{k_1}\cdot\overrightarrow{\sigma}\otimes\overrightarrow{k_2}\cdot\overrightarrow{\sigma}$. The $4\times4$ matrix form of this observable reads
                        \begin{widetext}\begin{eqnarray*}\left(
                                     \begin{array}{cccc}
                                       k_{1z}  k_{2z}& k_{1z} (k_{2x}-k_{2y} i) & k_{2z} (k_{1x}-k_{1y} i) & (k_{1x}-k_{1y} i)(k_{2x}-k_{2y} i) \\
                                       k_{1z} (k_{2x}+k_{2y} i) & -k_{1z}  k_{2z}& (k_{1x}-k_{1y} i)(k_{2x}+k_{2y} i) &  - k_{2z}(k_{1x}-k_{1y} i) \\
                                       k_{2z} (k_{1x}+k_{1y} i)&  (k_{1x}+k_{1y} i) (k_{2x}-k_{2y} i) & -k_{1z}  k_{2z} &-k_{1z} (k_{2x}-k_{2y} i)\\
                                       (k_{1x}+k_{1y} i) (k_{2x}+k_{2y} i) & -k_{2z}(k_{1x}+k_{1y} i) & -k_{1z} (k_{2x}+k_{2y} i) & k_{1z}  k_{2z} \\
                                     \end{array}
                                   \right)
      \end{eqnarray*}\end{widetext}
      For the two-qubit system the measuring of observable $\widehat{O}$ is connected with the measuring spin projection of the first spin on the direction given by the vector $\overrightarrow{k_1}$ and for the second spin projection on the direction given by the vector $\overrightarrow{k_2}$. In fact, this observable is determined by two commuting observables $\widehat{O}_1=\overrightarrow{k_1}\cdot\overrightarrow{\sigma}\otimes\widehat{1}_2$ and $\widehat{O}_2=\widehat{1}_1\otimes\overrightarrow{k_2}\cdot\overrightarrow{\sigma}$ which have matrix representations
            \begin{eqnarray*}&&O_1=\left(
                                               \begin{array}{cccc}
                                                  k_{1z}  & 0 & k_{1x}-k_{1y} i & 0 \\
                                                 0 &  k_{1z} & 0 & k_{1x}-k_{1y} i \\
                                                 k_{1x}+k_{1y} i &0 & -k_{1z} & 0 \\
                                                 0 & k_{1x}+k_{1y} i  & 0 & -k_{1z} \\
                                               \end{array}
                                             \right)
            ,
             \end{eqnarray*}
             \begin{eqnarray*}&&O_2=\left(
                                               \begin{array}{cccc}
                                                  k_{2z}  & k_{2x}-k_{2y} i &0  & 0 \\
                                                 k_{2x}+k_{2y} i &  -k_{2z} & 0 & 0\\
                                                 0 &0 &  k_{2z}  &k_{2x}-k_{2y} i \\
                                                 0 &0  & k_{2x}-k_{2y} i &  -k_{2z}  \\
                                               \end{array}
                                             \right).
            \end{eqnarray*}
      It is obvious that $\widehat{O}=\widehat{O}_1\cdot\widehat{O}_2$ holds. In this sense the commuting observables $\widehat{O}_1$ and $\widehat{O}_2$ can be measured simultaneously and the measurement of the correlation function $E(\overrightarrow{k}_1,\overrightarrow{k}_2)$
       or the measurement of the observable $\widehat{O}$ is reduced to the measuring the observables $\widehat{O}_1$ and $\widehat{O}_2$ for the two qubits or the two-level atom.
       \par The physical meaning of the correlation function is the following. The correlation function for the two qubits is used to discuss the Bells inequalities \cite{Bell,Horn} and their violation is detected, e.g., in \cite{Aspect}.
       In the case of the Bell inequalities the correlation function $E(\overrightarrow{k}_1,\overrightarrow{k}_2)$ is considered for four pairs of directions $\overrightarrow{k}_1$ and $\overrightarrow{k}_1$, namely $(\overrightarrow{k}_1,\overrightarrow{k}_2)=\{(\overrightarrow{a},\overrightarrow{b});(\overrightarrow{a},\overrightarrow{c});(\overrightarrow{d},\overrightarrow{b});(\overrightarrow{d},\overrightarrow{c})\}$.
       It is known that for the two qubits the Bell inequality reads as
       \begin{eqnarray*}&&|E(\overrightarrow{a},\overrightarrow{b})+E(\overrightarrow{a},\overrightarrow{c})+E(\overrightarrow{d},\overrightarrow{b})-E(\overrightarrow{d},\overrightarrow{c})|\leq2  \end{eqnarray*}
       for separable states and
        \begin{eqnarray*}&&|E(\overrightarrow{a},\overrightarrow{b})+E(\overrightarrow{a},\overrightarrow{c})+E(\overrightarrow{d},\overrightarrow{b})-E(\overrightarrow{d},\overrightarrow{c})|\leq2\sqrt{2}  \end{eqnarray*}
        holds for the entangled states.
         In \cite{Zukowski} the steering inequality for the two-qubit state is based on the maxima of the correlation function \eqref{4} given in the following form
 \begin{eqnarray*}E(\overrightarrow{k}_1,\overrightarrow{k}_2) &=&\sum\limits_{i,j=1}^{3}T_{ij}k_{1i}k_{2j},
    \end{eqnarray*}
 where  $T_{ij}$ are components of the correlation matrix. If the bypartite state  is non-steerable, then the following inequality is fulfilled
   \begin{eqnarray}\label{5}\max_{\overrightarrow{k}_1,\overrightarrow{k}_2}(E(\overrightarrow{k}_1,\overrightarrow{k}_2))&\geq&\frac{2}{3}\sum\limits_{i,j=1}^{3}T_{ij}.
       \end{eqnarray}
       One can see that the Bell inequalities and the steering phenomenon reflect different aspects of the quantum correlations in two qubit system.
       \par To formulate the steering phenomenon for the four level atom or for the spin $j=3/2$ states we apply the same formalism used for two-qubit system states. We introduce two
       observables $\widehat{\mathcal{O}}_1$ and $\widehat{\mathcal{O}}_2$ (i.e Hermitian operators) for the single qudit system. The observables $\widehat{\mathcal{O}}_1$ and $\widehat{\mathcal{O}}_2$ (operators) act in the Hilbert space of the states of the four-level atom ($j=3/2$ qudit).
       The matrix form of the observables $\widehat{\mathcal{O}}_1$, $\widehat{\mathcal{O}}_2$ in the basis $|3/2, 3/2>$, $ |3/2, 1/2>$, $|3/2, -1/2>$, $|3/2, -3/2>$ of the spin system is identical to $\widehat{O}_1$ and $\widehat{O}_2$.
       \par Then we introduce the observable $\widehat{\mathcal{O}}=\widehat{\mathcal{O}}_1\cdot \widehat{\mathcal{O}}_2$ which is the Hermitian operator acting in the Hilbert space $\mathcal{H}$ which is not considered as a product, i.e. $\mathcal{H}\neq\mathcal{H}_1\otimes\mathcal{H}_2$. This product form was used for the system of the two qubits. Nevertheless due to postulates of qunatum mechanics any observable (the Hermitian operator) can be measured. We suggest to introduce the steering notion for the single qudit (e.g., the spin $j=3/2$) to be based on the measuring of the correlation function $\mathcal{E}(\overrightarrow{k}_1,\overrightarrow{k}_2)$ which corresponds to the measuring of the observable $\widehat{\mathcal{O}}$ in states of the single qudit.  It is the main tool to translate the steering properties known for the composite systems (two qubits) to the
       noncomposite systems as well as a single qudit (four-level atom).
       \par Hence, analogically to \eqref{14} we can write the correlation function of the single qudit state as
          \begin{eqnarray}\label{15}\mathcal{E}(\overrightarrow{k}_1,\overrightarrow{k}_2)&=&\int W_{k_1\sigma\otimes k_2\sigma}(\mathbf{y})
         W_{\rho}^d(\mathbf{y})d\mathbf{y},
   \end{eqnarray}
   where the tomograms are defined by \eqref{6}.
   \par Using the intertwining kernel \eqref{9} we can deduce that the correlation functions \eqref{14} and \eqref{15} are mathematically completely equivalent.
             \begin{widetext}\begin{eqnarray}\label{19}E(\overrightarrow{k}_1,\overrightarrow{k}_2) &=&\sum\limits_{m_1,m_2=-1/2}^{1/2}\int\int d\vec{n_1}d\vec{n_2}\sum\limits_{m=-3/2}^{3/2}\int W_\rho(m,\vec{n})W^{d}_{k_1\sigma\otimes k_2\sigma}(m,\vec{n})K_{12}K_{21}d\vec{n}=\mathcal{E}(\overrightarrow{k}_1,\overrightarrow{k}_2).
    \end{eqnarray}
             \end{widetext}
    Here we use that $K_{12}K_{21}=1$ holds.
   \subsection{The Example}
   If the Werner density matrix \eqref{10_1} describes the two-qubit state then the correlation function \eqref{14} is
    \begin{eqnarray}\label{18}E(\overrightarrow{k_1},\overrightarrow{k_2})&=&k_{1z}k_{2z}p.
   \end{eqnarray}
   The correlation tensor \eqref{5} is the diagonal matrix with entries $p(1-1 \quad 1)$ and the maximum value of the correlation function \eqref{18} is equal to $p$ in the domain $0<p<1/3$ and to $-p$ in $-1/3<p\leq0$. Hence, the inequality \eqref{5} is fulfilled if
$0<p<1/2$ holds. Since we are interested only in the entangled states, the parameter domain $1/3<p<1/2$ corresponds to the steerable Werner state.
\par If the Werner density matrix \eqref{10_1} describes the single qudit $j=3/2$ state it is straightforward to verify that the correlation function \eqref{14} has the form \eqref{18} that coincides with \eqref{19}.
\section{Summary}
To resume we formulate the main result of our work. We have shown that the quantum correlations reflected by the phenomenon of the quantum steering available in the two-qubit system
take place also in the single qudit $j=3/2$. We demonstrate the inequalities for the correlation function detecting the presence of the steering not only for the two-qubit states but also for the single qudit $j=3/2$ state. The physical meaning of these correlations is different from the case  of two qubits. The observables to be measured for the obtained correlation
being mathematically completely equivalent  to observables measured in the experiment with two qubits are different for the single qudit. The results are obtained by using the tomographic probability representation of the quantum states and the intertwining kernel related to  the two-qubit state tomogram and to the qudit $j=3/2$ tomogram is explicitly calculated.
The results are illustrated by the example of the Werner density matrix that can describe the two-qubit and the single qudit states.
The extension of the steering consideration for the other single qudits will be done in our future publication.


\bibliography{apssamp}

\end{document}